\documentstyle[12pt]{article}
\newcommand\mathC{\mkern1mu\raise2.2pt\hbox{$\scriptscriptstyle|$}
        {\mkern-7mu\rm C}} 
\newcommand{\mathR}{{\rm I\! R}}         

\newcommand\mapdown[1]{\Big\downarrow
                        \rlap{$\vcenter{\hbox{$\scriptstyle#1$}}$}}
\newcommand\mapright[1]{\smash{
        \mathop{\mbox{\large{$\longrightarrow$}}}\limits^{#1}}}
\newcommand\bundle[3]{\begin{array}[t]{c}
        {#1}\\ \mapdown{#2}\\ {#3}\end{array}}
\newcommand\bundlemap[2]{\begin{array}[t]{c}
\mapright{#1}\\
\phantom{\mapdown{}}\\\mapright{#2}\\\end{array}}

\begin{document}
\begin{titlepage}
\hspace{7truecm}Imperial/TP/01-02/20
\begin{center}
{\large\bf Some Reflections on the Status of Conventional Quantum
Theory when Applied to Quantum Gravity}
\end{center}

\vspace{0.8 truecm}

\begin{center}
        C.J.~Isham\footnote{email: c.isham@ic.ac.uk}\\[10pt]
        The Blackett Laboratory\\
        Imperial College of Science, Technology \& Medicine\\
        South Kensington\\
        London SW7 2BZ\\
\end{center}

\begin{center}
14 May 2002
\end{center}
All current approaches to quantum gravity employ essentially
standard quantum theory including, in particular, continuum
quantities such as the real or complex numbers. However, I
wish to argue that this may be fundamentally wrong in so far
as the use of these continuum quantities in standard quantum
theory can be traced back to certain {\em a priori\/}
assumptions about the nature of space and time: assumptions
that may be incompatible with the view of space and time
adopted by a quantum gravity theory. My conjecture is that in,
some yet to be determined sense, to each type of space-time
there is associated a corresponding type of quantum theory in
which continuum quantities do not necessarily appear, being
replaced with structures that are appropriate to the specific
space-time.

Topos theory then arises as a possible tool for `gluing'
together these different theories associated with the
different space-times. As a concrete example of the use of
topos ideas, I summarise recent work applying presheaf theory
to the Kochen-Specher theorem and the assignment of values to
physical quantities in a quantum theory.

\end{titlepage}
\section{Introduction}
The period in the late 1960's when I was a postgraduate
student at Imperial College saw rapid changes in theoretical
physics in response to data streaming from the world's
particle accelerators. One consequence was that the subject
matter of a student's PhD thesis sometimes changed
uncomfortably rapidly during the course of his or her studies.
As a result, there was a tendency for supervisors to assign a
provisional thesis title like ``Topics in elementary particle
physics''---a practice that was understandable, but which was
finally blocked by the University of London some years ago!

When asked to speak at this Symposium in honour of Stephen
Hawking, I adopted a similar tactic by choosing as the
provisional title ``Prima facie questions in quantum
gravity'', on the grounds that this would give maximum
flexibility when it came to actually write the talk. However,
in the event, I have chosen to focus on one single issue, and
the title of my lecture has been readjusted accordingly.

The question I wish to address is the extent to which ideas of
standard quantum theory are adequate for the formulation of a
quantum theory of gravity: in particular, in regard to (i) the
use of continuum quantities in the mathematical foundations of
quantum theory; and (ii) possible roles for topos theory. In
this context it should be emphasised that all the current
mainstream approaches to quantum gravity use standard quantum
theory in one form or another.

Of course, it is well understood that, at a {\em conceptual\/}
level, the standard interpretation of quantum theory is
inadequate when, for example,  applied to quantum cosmology.
Specifically, the lack of any external observer of the
universe `as a whole' throws into doubt the instrumentalism of
the Copenhagen interpretation; as does any attempt to
construct a quantum gravity theory with no background
space-time in which an `observer' could be placed. The extent
to which such reservations apply to quantum gravity away from
the cosmological regime is still debated, but in practice most
work on quantum gravity pays only lip service to these
conceptual issues.

However, what I have in mind are not conceptual issues {\em
per se\/} but rather certain {\em mathematical\/} ingredients
in the formalism of quantum theory that are invariably taken
for granted and yet which, I claim, implicitly assume certain
properties of space and time that may be fundamentally
incompatible with the spatio-temporal concepts needed for a
successful quantum gravity theory. An example of particular
interest is the use of the {\em continuum\/} (via the real or
complex numbers).

More generally, one can question the almost universal
assumption that spatio-temporal concepts are to be implemented
mathematically using standard point-set theory: this
notwithstanding the frequently-voiced objection that the
literal idea of a space, or time, point is physically
meaningless.  In fact, there exists something---namely topos
theory---that can replace set theory as the foundation of
mathematics, and which could arguably be a more appropriate
way of modelling spatio-temporal concepts in physical regimes
where quantum gravity effects are paramount. As we shall see,
topos theory is also relevant to questions concerning the
status of the continuum. These considerations have motivated
my focussing the lecture on two main areas: (i) the {\em a
priori\/} status of spatio-temporal concepts in quantum
theory, particularly in regard to the use of continua; and
(ii) certain possible roles for topos theory in theoretical
physics.

Considerations of this type are part of the general question
of the role of standard spatio-temporal concepts in a theory
of quantum gravity. In the current major quantum gravity
programmes, most of these concepts are inserted by hand as
part of the overall background structure of the theory. On the
other hand, there is a school of thought that maintains that
the standard ideas of space and time should `emerge' from the
theory only in some appropriate limit or physical regime; in
which case, a crucial question is whether the theory contains
{\em any\/} fundamental concepts/structures that can be be
broadly identified as `spatio-temporal', or if all such
concepts or structures are emergent in some way. One of the
attractions of the consistent-histories approach to quantum
theory (of which more later) is that it allows for the idea of
emergent structures in a natural way via the process of
coarse-graining: analogous to how thermodynamical concepts
arise from statistical physics when microscopic details of the
system are ignored.

{\em En passent\/}, one might ask what else could arise from
the fundamental theory in some appropriate physical limit.
This might include the entire mathematical formalism of
standard quantum theory, including its use of Hilbert spaces
defined over $\mathR$ or $\mathC$. Perhaps the conceptual
structure of standard quantum theory is also an emergent
structure: in particular, the special role for measurement,
and the use of probabilities that lie in the closed interval
$[0,1]$ of the real numbers. Certainly, there is no compelling
logical reason why whatever plays the role of standard quantum
theory at the Planck length---if, indeed, there is {\em any\/}
such theory---should possess all the features of the theory
that is known to work empirically only at atomic and nuclear
scales.

However, one of the key questions of interest in the present
paper is not how the standard ideas of space and time (and
probability) might {\em emerge\/} from a different formalism;
but rather how one might proceed to construct a quantum theory
{\em ab initio\/} in which whatever fundamental
spatio-temporal concepts are present are definitely not the
familiar continuum ones: for example, if one is given a finite
causal set as a background structure. The first step, and the
only one taken in the present paper, is to sound a cautionary
note by emphasising how strongly the continuum ideas of space
and time are implicitly embedded in the standard formulation
of quantum theory.

 The plan of the paper is as follows. In Section
\ref{Sec:APriori} there is a discussion of the role of
continuum concepts in the formulation of quantum theory in the
presence of a non-standard background (such as a causal set
\cite{BMLS87} \cite{Sorkin91}). The conclusion of the
discussion is that, in some appropriate sense, there may be a
different {\em type\/} of quantum theory for each type of
background space-time.

If this is indeed the case, the question then arises as to how
these different theories are to be `patched' together in a
true quantum gravity theory in which these backgrounds are
themselves subject to quantum effects. One possibility is the
use of {\em topos\/} theory. In Section \ref{Sec:AltConceptST}
we discuss other ways in which the standard notion of
space-time may change and, again, find a possible role for
topos theory.

Since topos theory is an important mathematical ingredient in
our considerations, one part of the subject---the theory of
presheaves---is introduced in Section
\ref{Sec:PreRelNotfroTop}. It is then shown in Section
\ref{Sec:PreProValQuaThe} how this can be used in a natural
way to illustrate certain key features (specifically the
Kochen-Specker theorem) of standard quantum theory. The main
physical idea here is a role for contextual, multi-valued
logic---an idea that in itself has many possible fruitful
applications in theoretical physics.

\section{The Danger of {\em A Priori Assumptions\/}}\label{Sec:APriori}
\subsection{The use of the real and complex numbers in quantum
theory} The use of the real and complex numbers is a basic
feature of all approaches to quantum theory: Hilbert spaces of
states, $C^*$-algebras of observables, quantum logic of
propositions, functional integral methods, etc., etc. These
number systems have a variety of relevant mathematical
properties, but the one of particular interest here is that
they are {\em continuua\/}, by which---in the present
context---is meant not only that $\mathR$ and $\mathC$ have
the appropriate cardinality, but also that they come equipped
with the familiar topology and differential structure that
makes them manifolds of real dimension one and two
respectively.

My concern is that the use of these numbers may be problematic
in the context of a quantum gravity theory whose underlying
notion of space and time is different from that of a smooth
manifold. The danger is that by imposing a continuum structure
in the quantum theory {\em a priori\/}, one may be creating a
theoretical system that is fundamentally unsuitable for the
incorporation of spatio-temporal concepts of a non-continuum
nature: this would be the theoretical-physics analogue of what
a philosopher might call a `category error'. For this reason,
it is important to consider carefully the origin, and role, in
standard quantum theory of this particular facet of the real
(and complex) numbers.

In general terms, the real numbers arise in three ways in
physical theories: (i) as the values of physical quantities;
(ii) to model space and time; and (iii) as the values of
probabilities. Our present task is to consider more precisely
the use real numbers in quantum theory in these terms.

As a first step, consider the simple example with which most
undergraduate courses on quantum theory begin: a
non-relativistic point particle moving in one dimension. The
state of the system at a time $t$ is represented by a wave
function $\psi_t(x)$, and we see at once that continuum
quantities are involved in three ways: (i) as the argument $x$
in the wave function; (ii) as the value of the wave function;
and (iii) as the time parameter $t$. Let us consider these in
turn.

\subsubsection{The $x$ in $\psi(x)$}
From one perspective, the $x$ in $\psi(x)$ arises because we
are starting with a classical theory and then `quantising' it.
In the present example, the classical configuration space $Q$
is identified with the real line because the system is a point
particle moving in (one-dimensional) physical space, and the
latter is modelled by the real numbers.

In general, the configuration space (if there is one) $Q$ for
a classical system is modelled mathematically by a
differentiable manifold, and the classical state space is the
co-tangent bundle $T^*Q$. The physical motivation for using a
manifold to represent $Q$ again reduces to the fact that we
represent physical space with a manifold. This is clearly so
for configurations that correspond to the position of the
centre-of-mass of an object in space, or its overall
orientation in space, but it also applies to internal degrees
of freedom of relative positions of constituent entities.

Thus, in assuming that the state space of a classical system
is a manifold of the form $T^*Q$ we are importing into the
classical theory a powerful {\em a priori\/} picture of
physical space: namely, that it is a differentiable
manifold\footnote{There may be cases where $\cal S$ is a
symplectic manifold that is not a cotangent bundle; for
example, ${\cal S}:=S^2$. However, I would argue that the
reason $\cal S$ is assumed to be a {\em manifold\/} is still
ultimately grounded in an {\em a priori\/} assumption about
the nature of physical space (and time).}. This then carries
across to the corresponding quantum theory. For example, if
`quantisation' is construed to mean defining the quantum
states to be cross-sections of some flat\footnote{The bundle
is chosen to be flat so that a covariant derivative of
sections can be defined without the need to introduce extra
local `connection' variables into the theory.} vector bundle
over $Q$, then the domain of these state functions is the
continuum space $Q$.

However, for this argument to have any force we need to
consider {\em why\/} quantisation is so defined, and this
takes us to the issue of the space in which the wave function
has its values.

\subsubsection{The value of the function $\psi(x)$---the role of
classical physical quantities.} In the example of the quantum
theory of a particle moving in one-dimension, the value of the
state function $\psi(x)$ is a complex number: so, once again,
a continuum concept arises. This particular one comes from two
different sources.

On the one hand, the operator $\hat x$ that represents the
position of the particle acts on the state function as
\begin{equation}
                (\hat x\psi)(x):=x\psi(x).
\end{equation}
More generally, for a system with a configuration manifold
$Q$, a classical physical configuration quantity corresponds
to a real-valued function $f:Q\rightarrow \mathR$, and this
function is represented in the quantum theory by the operator
\begin{equation}
                (\hat f\psi)(q):=f(q)\psi(q).
                \label{Def:hatf}
\end{equation}
on sections of the appropriate vector bundle.

Although this equation does not {\em prove\/} that $\psi(q)$
is $\mathC$-valued, it does show that, for each $q\in Q$, the
space in which $\psi(q)$ takes its values must be such that it
admits a {\em multiplication\/} operation by real numbers.

We see that this particular sources of the real numbers in
quantum theory comes from the assumption that {\em
classical\/} physical quantities are real-valued, which is
then translated into an analogous requirement on the quantum
variables.

A related feature is that in any quantum system the
eigenvector equation for a physical quantity $A$ is of the
form $\hat A |a\rangle = a|a\rangle$, where $a$ is a real
number. So the state space has to be such that its elements
can be multiplied by real numbers. Note that this applies even
to quantum physical quantities that have no classical
analogue: we still assume that their eigenvalues are real
numbers. Of course, for many quantities the set of all
eigenvalues will be a discrete subset of $\mathR$, but that
does not detract from the point being made here.

 It is thus pertinent to ask {\em why\/} physical
quantities---classical or quantum---are taken to be
real-valued. Many will doubtless say that the answer is
obvious, or that it is even part of the definition of a
physical quantity, but I would challenge these assertions as
being over-hasty.

One reason why the values of physical quantities are assumed
to be real numbers is undoubtedly the operational one
that---at least, in the pre-digital age---physical quantities
are ultimately measured with rulers and pointers, and so it is
the assumed continuum nature of physical space that comes into
play.

However, it is by no means obvious that physical quantities
should necessarily be real-valued in, for example, a quantum
gravity theory in which it is not appropriate to think of
space as a smooth manifold, and where, therefore, there is no
place for operational considerations that presuppose a
continuum nature for space and/or time.

Of course, it is a totally open question as to what should
replace $\mathR$ as the value space of a physical quantity in
these circumstances---it could be something as obvious as a
finite number field, but it could also be something far more
radical. In any event, a key role in deciding this issue
should be played by any underlying spatio-temporal concepts
(albeit, non-standard) that are present.

\subsubsection{The value of the function $\psi(x)$
---the role of probability.}
A different source of the $\mathC$-valued nature of the
wave-function is its probabilistic interpretation. Of course,
this extends outside simple wave mechanics, with the general
quantum-theory result that if $\hat E(A\in\Delta)$ is the
spectral projector onto the eigenspace of $\hat A$ with
eigenvalues in the (Borel) set $\Delta\subset\mathR$ then, if
the (normalised) state is $\psi$, the probability that the
proposition ``The physical quantity $A$ lies in $\Delta$'' is
true is\footnote{Of course, in the standard Copenhagen
interpretation of quantum theory it would be more appropriate
to say that the proposition represented by the spectral
projector $\hat E(A\in\Delta)$ is ``If a measurement is made
of $A$, then the value will be found to lie in $\Delta$.}
\begin{equation}
        {\rm Prob}(A\in\Delta;\psi)=\langle\psi, \hat
        E(A\in\Delta)\rangle.       \label{Eqn:ProbAinDelta}
\end{equation}

From our present perspective, the key point  is that the
assumption that probabilities should lie in the interval
$[0,1]$ of the real numbers requires the field over which the
Hilbert state space is defined to be such as to accommodate
this assumption via the right hand side of Eq.\
(\ref{Eqn:ProbAinDelta}). So this is yet another source of the
use of continuum quantities in the mathematical formulation of
quantum theory.

In the context of standard physics, it is clear why
probabilities are required to lie in the interval $[0,1]$. As
physicists, we most commonly employ a relative-frequency
interpretation of probability in which an experiment is
repeated a large number, say $N$, times, and the probability
associated with a particular result is then defined to be the
ratio $N_i/N$, where $N_i$ is the number of experiments in
which that result was obtained. The rational numbers $N_i/N$
necessarily lie between $0$ and $1$, and if we take the limit
as $N\rightarrow\infty$, as is appropriate for a hypothetical
`infinite ensemble', we get real numbers in the closed
interval $[0,1]$.

Although the relative-frequency interpretation of probability
may seem natural in standard physics,  it is not meaningful in
situations where there is no classical spatio-temporal
background in which observations could be made; or, if there
is a background, it is such that there is no meaningful
analogue of the relative-frequencies interpretation adapted to
that background.

Under such circumstances it might be more natural to follow
Aristotle, Heisenberg and Popper in adopting a {\em
propensity\/} interpretation of probability, perhaps within
the context of a `post-Everett' form of quantum theory, such
as consistent-histories theory.

However, if probability is viewed in this more realist way,
there is no overwhelming reason for assigning its values to be
real numbers lying in the interval $[0,1]$. The minimal
requirement is presumably only that the value space should be
a partially ordered set $({\cal V},\leq)$ so that it makes
sense to say that certain events are more, or less, probable
(in the sense of the partial-ordering operation $\leq$) than
others.

Note that this allows for the possibility of pairs of events
whose propensities are incomparable: {\em i.e.}, the
probability value-space $\cal V$ may not be a {\em
totally}-ordered set. We would, however, expect there to be a
unit element $1\in\cal V$, corresponding to the probability of
an event that is certain to happen (or the proposition that is
identically true), and with $p \leq 1$ for all $p\in\cal V$.
Similarly, there should be a null element $0\in\cal V$,
corresponding to the probability of an event that is certain
not to happen (or the proposition that is identically false),
and with $0\leq p$ for all $p \in\cal V$.

It also seems natural to require that $\cal V$ has some
`semi-additive' structure so that the probability of two
disjoint events is the `sum' of the probabilities of the
individual events. At the very least, if $P$ is any
proposition and $\neg P$ is its negation, we would expect the
probability of $P\lor\neg P$ to be the unit\footnote{This
would not be so if for some reason the quantum propositions
obeyed an intuitionistic logic (see later) where the principle
of excluded middle does not necessarily apply.} $1\in\cal P$,
and equal to the `sum' of the probabilities of $P$ and $\neg
P$.

Of course, it is an open question as to what precise
mathematical structure should be used as the value-space for
probabilities in the absence of any classical spatio-temporal
background; or, indeed, what it should be in the presence of a
non-standard background such as a causal set. But the key
point is that there is no fundamental reason why this
value-space has to involve the real numbers; and the form of
quantum theory in such a situation should reflect this fact.

\subsubsection{The $t$ in $\psi_t(x)$}
The time-parameter $t$ in the wave-function is taken directly
from the corresponding parameter in classical,
non-relativistic physics. It is Newtonian time, and as such it
is part of the background structure of standard Newtonian
physics. It is represented by a real number: indeed, the full
manifold structure of $\mathR$ (and of the classical state
space) is invoked when defining the differential equations of
motion of classical physics.

In relativistic physics, space and time are placed on a more
equal footing, with a background space-time manifold rather
just a background time. In special relativity, this background
manifold has the topological and differential structure of
$\mathR^4$, and is equipped with the, fixed, Minkowskian
metric tensor.

Things change considerably when we come to the space and time
of general relativity: indeed in the context of quantum
gravity, time is a difficult concept---in particular, there is
the well-known `problem of time' that affects all approaches
to quantum gravity in one way or another.

This problem was first explicitly encountered in the context
of the canonical approach to quantum gravity, whose central
feature is the constraint equations on the state vector $\Psi$
\begin{eqnarray}
        \hat{\cal H}_i(x)\Psi&=&0       \label{Eqn:HiPsi=0}\\
        \hat{\cal H}_\perp(x)\Psi&=&0   \label{Eqn:HperpPsi=0}
\end{eqnarray}
where ${\cal H}_i$ and ${\cal H}_\perp$ are constructed from
the metric tensor $g$ (and its conjugate variable) on an
underlying 3-manifold \cite{isham-rev2} \cite{kuch-prehled}.

Equation (\ref{Eqn:HiPsi=0}) simply asserts the invariance of
$\Psi$ under (small) spatial diffeomorphisms. However,
equation (\ref{Eqn:HperpPsi=0}) is more problematic. In the
representation in which $\Psi$ appears as a functional
$\Psi[g]$, eq.\ (\ref{Eqn:HperpPsi=0}) is the Wheeler-DeWitt
equation, and---in one approach to the `problem of time'---is
interpreted as a dynamical equation with respect to an
`internal' time variable that has to be constructed from the
metric tensor and its conjugate. It is always assumed that
this variable will be represented by a real number: indeed,
the internal time is usually sought from the perspective of
{\em classical\/} canonical general relativity---which is
bound to lead to a real quantity. So once again we see how
{\em a priori\/} assumptions about the nature of time can be
placed into the quantum theory from the outset. (Of course,
the canonical approach already comes with an explicit
background {\em spatial\/} manifold.)

\subsection{Space-time dependent quantum theory} The main conclusion
I wish to draw from the discussion above is that a number of
{\em a priori\/} assumptions about the nature of space and
time are present in the mathematical formalism of standard
quantum theory, and it may therefore be necessary to seek a
major restructuring of this formalism in situations where the
underlying spatio-temporal concepts (if there are any at all)
are different from the standard ones which are represented
mathematically with the aid of differential geometry.

A good example would be to consider from scratch how to
construct a quantum theory when space-time is a finite causal
set: either a single such---which then forms a fixed, but
non-standard, spatio-temporal background---or else a
collection of such sets in the context of a type of quantum
gravity theory. In the case of a fixed background, this new
quantum formalism should be adapted to the precise structure
of the background, and can be expected to involve a
substantial departure from the standard formalism:
particularly in regard to the use of real numbers as the
values of physical quantities and probabilities.

The fundamental emphasis in a causal set is on a space-time
structure as a single unit, rather than separate space and
time structures, and this suggests strongly that it would be
better to start {\em ab initio\/} with a {\em history\/}
theory rather than one in which some type of `temporal
slicing' is introduced. It should be emphasised that the
path-integral approach to standard quantum theory is {\em
not\/} a history theory in the way the phrase is being used
here. Indeed, a path integral generates transition amplitudes
between canonical states, which implicitly requires some type
of time slicing. In fact, the only genuine `history' theory I
know that can handle space-time stuctures as integral entities
is the consistent-histories formalism of Griffiths
\cite{Gri84}, Omnes \cite{Omn88} and Gell-Mann and Hartle
\cite{GH90}.

Thus an instructive research programme would be to develop a
version of consistent-history quantum theory that is
appropriate for a background causal set. In this context, a
particularly useful approach could be the Gell-Mann and Hartle
method as axiomatised in the language of quantum temporal
logic by Isham \cite{Isham94}, and Isham and Linden
\cite{IL94}, together with the completely new perspective on
the role of time introduced by Savvidou \cite{Sav99a}. Here
one has an orthoalgebra $\cal UP$ of propositions about the
history of the system, and a space $\cal D$ of `decoherence
functions' that are maps $d:{\cal UP}\times{\cal
UP}\rightarrow\mathC$ and which encode both the dynamics and
the initial conditions. From a physical perspective, if a
proposition $\alpha\in\cal UP$ belongs to a consistent set,
then $d(\alpha,\alpha)$ is interpreted as the probability that
$\alpha$ is true in the context of that consistent set.

It follows from the discussion above, that if there is a
background causal set the quantum history formalism should be
such that this structure is reflected in (i) the choice of the
space $\cal UP$ of propositions about the `universe'; and (ii)
the choice of the space in which decoherence functions take
their values, with an associated change in the mathematical
representation of probability.

Finally, if it is indeed the case that, in some sense, to each
background space-time there is associated a corresponding type
of quantum theory, then the question arises as to how these
different theories can be `patched' together to give a quantum
space-time theory in which the different backgrounds are
themselves the subject of quantum effects. One possibility is
the use of {\em topos\/} theory: in particular, the theory of
presheaves which provides a powerful way of handling
situations where there is a space of `contexts' with respect
to which individual structures are associated. For example, a
context could be a causal set.

Topos theory is of potential interest in theoretical physics
in a number of ways, and it will recur in much of what
follows. For this reason, an introduction to some of the basic
ideas in given in Section \ref{Sec:PreRelNotfroTop}.

\section{Alternative Conceptions of Spacetime}
\label{Sec:AltConceptST}
\subsection{Points or Regions?}
Doubts about the use of the continuum in present-day physical
theories prompts one to consider more general alternative
conceptions of space and time. We turn now to briefly sketch
two such, both of which involve topos theory, and which raise
the even more iconoclastic idea that the use of set theory
itself may be inappropriate for modelling space and time in
the context of quantum gravity.

\subsubsection{From points to regions}
\label{SubSec:Loc} In standard general relativity---and,
indeed, in all classical physics---space (and similarly time)
is modelled by a set, and the elements of that set correspond
to points in space. However, it is often claimed that the
notion of a spatial (or temporal) point has no real physical
meaning, and this motivates trying to construct a theory in
which `regions' are the primary concept. In such a
theory,`points'---if they exist at all---would play a
secondary role in which they are determined in some way by the
regions (rather than regions being collections of points, as
in standard set theory).

In fact, there are axiom systems for regions, some of whose
models do not contain anything corresponding to points of
which the regions are composed. As an example, consider a
topological space $X$. The family of all open sets has the
algebraic operations of conjunction, disjunction and negation
defined by $O_1\land O_2:=O_1\cap O_2$; $O_1\lor O_2:=O_1\cup
O_2$; and $\neg O:={\rm int}(X-O)$ respectively; and with
these operations, the open sets form a complete Heyting
algebra, also known as a {\em locale\/}. Here, a Heyting
algebra $H$ is defined to be a distributive lattice, with null
and unit elements, that is {\em relatively complemented\/},
which means that to any pair $S_1,S_2$ in $H$, there exists an
element $S_1\Rightarrow S_2$ of $H$ with the property that,
for all $S\in H$, we have $S\leq (S_1\Rightarrow S_2)\mbox{ if
and only if $S\land S_1\leq S_2$}$.

Heyting algebras are thus a generalization of Boolean
algebras. In particular, they need not obey the law of
excluded middle, and so provide natural algebraic structures
for intuitionistic logic. A Heyting algebra is said to be {\em
complete\/} if every family of elements has a least upper
bound. Thus, when partially ordered by set-inclusion, the open
sets of any topological space form a Heyting algebra. This
algebra is complete since arbitrary unions of open sets are
open, and the disjunction of an arbitrary family of open sets
can be defined as the interior of their intersection.

However, it transpires that not every locale is isomorphic to
the Heyting algebra of open sets of some topological space;
and in this sense, the theory of regions given by the
definition of a locale is a generalisation of the idea of a
topological space that allows regions that are not composed of
underlying points. This might be an interesting alternative to
standard topology for modelling space-time in the context of
quantum gravity.

A far-reaching generalisation of this idea is given by topos
theory. As we shall see in Section \ref{Sec:PreRelNotfroTop},
in any topos the idea of a `subobject' is the analogue of the
set-theoretic notion of a subset of a given set; and for any
object $X$ in a topos, the family of subobjects of $X$ is a
Heyting algebra, and hence another possible model for the
regions of space-time.

\subsection{Synthetic Differential Geometry}
\label{SubSec:SynDifGeo} Recent decades have seen a revival of
the idea of infinitesimals: nilpotent real numbers $d$ such
that $d^2 = 0$. At first sight this seems nonsensical (apart
from the trivial case $d = 0$) but it turns out that sense
{\em can\/} be made of this, and in two different ways.

In the first approach, called `non-standard analysis', every
infinitesimal has a reciprocal, so that there are different
infinite numbers corresponding to the different
infinitesimals. There were attempts in the 1970s to apply this
idea to quantum field theory: in particular, it was shown how
the different orders of ultra-violet divergences correspond to
different types of infinite number in the sense of
non-standard analysis \cite{Far75}.

In the second approach, there are infinitesimals but without
the corresponding infinite numbers. This is possible provided
we work within the context of a {\em topos\/} rather than
normal set theory: for example, a careful study of the proof
that the only real number $d$ such that $d^2=0$ is $0$, shows
that it involves the principle of excluded middle, which in
general does not hold in the intuitionistic logic of a topos
\cite{Lav96}.

This approach is known as `synthetic differential geometry'
(SDG), and it is intriguing to see if our familiar physical
theories can be rewritten using this structure. For example,
Fearns has recently shown how some of the features of standard
quantum theory can be expressed in this way \cite{Fearns02}
(see also \cite{Nishimura97}).

Of even greater importance, however, is the possibility that
there may be regimes in physics, in particular involving
quantum space-time structures, where SDG is {\em more\/}
appropriate than the standard approach.

One such possibility is suggested by the `History Projection
Operator'  approach to consistent histories in which there are
copies of the standard canonical commutation relations at each
moment of time. For example, for a particle moving in one
dimension we have the history algebra \cite{ILSS98}
\begin{eqnarray}
[x_{t},x_{t'}]&=&0 \label{CTHAxx} \\
{[}p_{t},p_{t'}]&=&0 \label{CTHApp} \\
{[}x_{t},p_{t'}]&=&i\hbar\delta(t'-t) \label{CTHAxp}
\end{eqnarray}
where the label $t$ on the (Schr\"odinger picture) operators
$\hat x_t$ and $\hat p_t$ refers to the time at which
propositions about the system are asserted---the time of
`temporal logic'.

A major advance in the HPO formalism took place when time was
introduced by Savvidou in a completely new way \cite{Sav99a}
\cite {Sav99b}. It was realised that it is natural to consider
time in a two-fold manner: the `time of being'---the time at
which events `happen' (the time label $t$ in Eqs.\
(\ref{CTHAxx})--(\ref{CTHAxp}) can be regarded as such), and
the `time of becoming'---the time of dynamical change,
represented by a time label $s$. This {\em second\/} time
appears in the history analogue $\hat x_t(s)$ of the
Heisenberg picture, which is defined as $\hat
x_t(s):=e^{is\hat H/\hbar}\hat x_t e^{-is\hat H/\hbar}$ where
$\hat H:=\int dt\hat H_t$ is the history quantity that
represents the time average of the energy of the system. The
notion of time evolution is now recovered for the
time-averaged physical quantities, for example $\hat
x_f(s):=e^{is\hat H/\hbar}\hat x_f e^{-is\hat H/\hbar}$ where
$f(t)$ is a smearing function.

Associated with these two manifestations of the concept of
time are two types of time transformation: the `external'
translation $\hat x_t(s)\mapsto\hat x_{t+t'}(s)$; and the
`internal' translation $\hat x_t(s)\mapsto\hat x_t(s+s')$. The
external time translation is generated by the `Louiville'
operator \cite{Sav99a} $\hat V:=\int dt\,\hat p_t {d\hat
x_t\over dt}$ whereas the internal time translation is
generated by the time-averaged energy operator $\hat H$.

More importantly, it was shown in \cite{Sav99a} that the
generator of time translation in the HPO theory is the
`action' operator $\hat S$ defined as
\begin{equation}
\hat S:=\int dt\,\hat p_t {d\hat x_t\over dt} - \hat H
        = \hat V-\hat H.
\end{equation}
Hence the action operator is the generator of {\em both\/}
types of time translation: $\hat x_t(s)\mapsto\hat
x_{t+t'}(s+s')$. It is a striking result that in the HPO
theory the quantum analogue of the classical action functional
is an actual operator in the formalism, and is the generator
of time translations.

In the context of SDG, it is the view of Savvidou (with which
I agree) that the infinitesimals of SDG are particularly well
adapted to describe transformations in the external
time-parameter.

If true, this has significant implications for the
construction of a history theory of quantum gravity. In
particular, in the context of general relativity, Savvidou has
shown that the analogue of the Liouville transformations is
the full space-time diffeomorphism group \cite{SavGR1}. Thus
the intriguing possibility arises that, in a history version
of general relativity, there may be a natural role for SDG,
and hence for topos theory, in implementing the actions of
this fundamental group.

\section{Presheaves and Related Notions from Topos Theory}
\label{Sec:PreRelNotfroTop} From now on we shall concentrate
on topos theory itself, culminating in a particular
application to standard quantum theory.

There are various approaches to the notion of a topos but the
focus here will be on one that emphasises the underlying
logical structure. To keep the discussion simple, we will not
develop the full definition of a topos but will concentrate on
the role of a `subobject classifier'. This involves a
generalization of the set-theoretic idea of a characteristic
function, and has a particularly interesting logical structure
in the kind of topos to which the discussion in Section
\ref{Sec:PreProValQuaThe} is confined: namely, a topos of
presheaves \cite{Gol84} \cite{MM92}.

A topos is a type of category that behaves much like the
category of sets $\rm Set$.\footnote{Recall that a category
consists of a collection of {\em objects\/} and a collection
of {\em arrows\/} (or {\em morphisms\/}), with the following
three properties: (1) Each arrow $f$ is associated with a pair
of objects, known as its {\em domain\/} (dom $f$) and the {\em
codomain\/} (cod $f$), and is written in the form
$f:B\rightarrow A$ where $B ={\rm dom} f$ and $A={\rm cod} f$;
(2) Given two arrows $f:B\rightarrow A$ and $g:C\rightarrow B$
(so that the codomain of $g$ is equal to the domain of $f$),
there is a composite arrow $f\circ g:C\rightarrow A$, and this
composition of arrows obeys the associative law; and (3) Each
object $A$ has an identity arrow, ${\rm id}_A:A\rightarrow A$,
with the properties that for all $f:B\rightarrow A$ and all
$g:A\rightarrow C$, ${\rm id}_A\circ f = f$ and $g\circ {\rm
id}_A = g$.} In the category $\rm Set$, the objects are sets
and the arrows/morphisms are ordinary functions between them
(set-maps). In many other categories, the objects are sets
equipped with some type of additional structure, and the
arrows are functions that preserve this structure. An example
is the category of groups, where an object is a group, and an
arrow $f:G_1\rightarrow G_2$ is a group homomorphism from
$G_1$ to $G_2$. However, a category need not have `structured
sets' as its objects. An example is given by any
partially-ordered set (`poset') $\cal P$. It can be regarded
as a category in which (i) the objects are the elements of
$\cal P$; and (ii) if $p,q\in\cal P$, an arrow from $p$ to $q$
is defined to exist if, and only if, $p\leq q$ in the poset
structure. Thus, in a poset regarded as a category, there is
at most one arrow between any pair of objects $p,q\in\cal P$.

In any category, an object $T$ is called {\em a terminal\/}
(resp.\ {\em initial\/}) object if for every object $A$ there
is exactly one arrow $f:A\rightarrow T$ (resp.\
$f:T\rightarrow A$). Any two terminal (resp.\ initial) objects
are isomorphic\footnote{Two objects $A$ and $B$ in a category
are said to be {\em isomorphic\/} if there exists arrows
$f:A\rightarrow B$ and $g:B\rightarrow A$ such that $f\circ
g={\rm id}_B$ and $g\circ f={\rm id}_A$}. So we can fix on one
such object and write `the' terminal (resp.\ initial) object
as ${\bf 1}$ (resp. ${\bf 0}$). An arrow ${\bf 1}\rightarrow
A$ is called a {\em point\/}, or {\em global element\/}, of
$A$. For example, applying these definitions to the category
of sets, we see that (i) each singleton set is a terminal
object; (ii) the empty set $\emptyset$ is initial; and (iii)
the points of $A$ are in one-to-one correspondence with the
elements of $A$ (in the usual sense of the word `element' of a
set).

\subsection{Toposes and Subobject Classifiers}
\label{SubSec:TopSubCla} We turn now to the very special kind of
category called a `topos', concentrating on the requirement that a
topos contains a generalization of the set-theoretic concept of a
characteristic function.

Recall that for any set $X$, and any subset $A \subseteq X$, there
is a characteristic function $\chi_A:X\rightarrow\{0,1\}$, with
$\chi_{A}(x) = 1$ or 0 according as $x \in A$ or $x \notin A$. One
can think of $\{0,1\}$ as truth-values, with $\chi_A$ classifying
the various $x\in X$ in response to question ``Is $x$ an element
of $A$?''. Furthermore, $\{0,1\}$ is itself a set---{\em i.e.\/}
an object in the category $\rm Set$---and for each $A, X$ with $A
\subseteq X$, $\chi_A$ is an arrow from $X$ to $\{0,1\}$.

These concepts extend to a general category as follows.
\begin{enumerate}
\item A `subobject' is the analogue of the set-theoretic idea of a subset. More precisely, one
generalizes the idea that a subset $A$ of $X$ has a preferred
injective ({\em i.e.\/}, one-to-one) map $A \rightarrow X$
sending $x \in A$ to $x \in X$. The categorial analogue of an
injective map is called a `monic arrow', and a subobject of
any object $X$ in a category is defined to be a monic arrow
with codomain $X$.

\item Any topos is required to have an analogue, written $\Omega$,
of the set $\{0,1\}$ of truth-values; in particular, $\Omega$
is an {\em object\/} in the topos. Furthermore, there is a
one-to-one correspondence between subobjects of an object $X$,
and arrows from $X$ to $\Omega$.

\item In a topos, $\Omega$ acts as an object of generalized truth-values,
just as $\{0,1\}$ does in set-theory (though $\Omega$
typically has more than two global elements). Moreover,
$\Omega$ has a natural logical structure. More precisely,
$\Omega$ has the internal structure of a Heyting algebra
object: the algebraic structure appropriate for intuitionistic
logic, mentioned in Section \ref{SubSec:Loc}. In addition, the
collection of subobjects of any given object $X$ in a topos is
a complete Heyting algebra.
\end{enumerate}

\subsection{Toposes of Presheaves}
\label{Subsec:TopPre} In preparation for the application to
quantum theory discussed in Section \ref{Sec:PreProValQuaThe},
we turn now to the theory of presheaves\footnote{More
precisely, the theory of presheaves on an arbitrary `small'
category $\cal C$ (the qualification `small' means that the
collection of objects in $\cal C$ is a genuine set, as is the
collection of all arrows in $\cal C$).}.

First recall that a `functor' between a pair of categories
$\cal C$ and $\cal D$ is a arrow-preserving function from one
category to the other. More precisely, a {\em covariant
functor\/} $\bf F$ from a category $\cal C$ to a category
$\cal D$ is a function that assigns (i) to each $\cal
C$-object $A$, a $\cal D$-object ${\bf F}(A)$; and (ii) to
each $\cal C$-arrow $f:B\rightarrow A$, a $\cal D$-arrow ${\bf
F}(f):{\bf F}(B)\rightarrow {\bf F}(A)$ such that ${\bf
F}({\rm id}_A)={\rm id}_{{\bf F}(A)}$. These assignments are
such that if $g:C\rightarrow B$, and $f:B\rightarrow A$ then
${\bf F}(f\circ g)={\bf F}(f)\circ {\bf F}(g)$.

A {\em presheaf\/} (or {\em varying set\/}) on the category
$\cal C$ is defined to be a covariant functor $\bf X$ from the
category $\cal C$ to the category of sets. We want to make the
collection of presheaves on $\cal C$ into a category, and so
it is necessary to define what is meant by an `arrow' between
two presheaves $\bf X$ and $\bf Y$. This is defined to be a
{\em natural transformation\/} $N:{\bf X}\rightarrow{\bf Y}$,
which is a family of maps (the {\em components\/} of $N$)
$N_A:{\bf X}(A)\rightarrow{\bf Y}(A)$, where $A$ an object in
$\cal C$, such that if $f:A\rightarrow B$ is an arrow in $\cal
C$, then the composite map ${\bf X}(A)
\stackrel{N_A}\longrightarrow{\bf Y}(A)\stackrel{{\bf Y}(f)}
\longrightarrow{\bf Y}(B)$ is equal to ${\bf X}(A)
\stackrel{{\bf X}(f)}\longrightarrow{\bf X}(B)\stackrel{N_B}
\longrightarrow {\bf Y}(B)$, as shown in the commutative
diagram
\begin{equation}
\bundle{{\bf X}(A)}{N_A}{{\bf Y}(A)} \bundlemap{{\bf X}(f)}{{\bf
Y}(f)} \bundle{{\bf X}(B)}{N_B}{{\bf Y}(B)}
\end{equation}

An object $\bf K$ is said to be a {\em subobject\/} of $\bf X$
if there is an arrow in the category of presheaves $i:{\bf
K}\rightarrow{\bf X}$ with the property that, for each $A$,
the component map $i_A:{\bf K}(A)\rightarrow{\bf X}(A)$ is a
subset embedding, {\em i.e.}, ${\bf K}(A)\subseteq {\bf
X}(A)$. Thus, if $f:A\rightarrow B$ is any arrow in $\cal C$,
we get the commutative diagram
\begin{equation}
\bundle{{\bf K}(A)}{}{{\bf X}(A)} \bundlemap{{\bf K}(f)}{{\bf
X}(f)} \bundle{{\bf K}(B)}{}{{\bf X}(B)} \label{subobject}
\end{equation}
where the vertical arrows are subset inclusions.

The category of presheaves on $\cal C$, ${\rm Set}^{{\cal
C}}$, forms a topos. We turn now to discussing the subobject
classifier of this particular topos.

\subsubsection{Sieves and the Subobject Classifier in a Topos of
Presheaves.} \label{SubSub:Sieves} A key concept in presheaf
theory---and something of particular importance for the quantum
theory application discussed later---is that of a `sieve', which
plays a central role in the construction of the subobject
classifier in the topos ${\rm Set}^{{\cal C}}$ of presheaves on a
category $\cal C$.

A {\em sieve\/} on an object $A$ in $\cal C$ is defined to be
a collection $S$ of arrows $f:A\rightarrow B$ in $\cal C$ with
the property that if $f:A\rightarrow B$ belongs to $S$, and if
$g:B\rightarrow C$ is any arrow, then $g\circ f:A\rightarrow
C$ also belongs to $S$. In the simple case where $\cal C$ is a
poset, a sieve on $p\in\cal C$ is any subset $S$ of $\cal C$
such that if $r\in S$ then (i) $p\leq r$, and (ii) $r'\in S$
for all $r\leq r'$. Thus a sieve is just an {\em upper\/} set
in the poset.

The presheaf ${\bf\Omega}:{\cal C}\rightarrow {\rm Set}$ is
now defined as follows. If $A$ is an object in $\cal C$, then
${\bf\Omega}(A)$ is defined to be the set of all sieves on
$A$; and if $f:A\rightarrow B$, then
${\bf\Omega}(f):{\bf\Omega}(A)\rightarrow{\bf\Omega}(B)$ is
defined as
\begin{equation}
{\bf\Omega}(f)(S):= \{h:B\rightarrow C\mid h\circ f\in S\}
                                \label{Def:Om(f)}
\end{equation}
for all $S\in{\bf\Omega}(A)$. Note that if $S$ is a sieve on
$A$, and if $f:A\rightarrow B$ belongs to $S$, then from the
defining property of a sieve
\begin{equation}
        {\bf\Omega}(f)(S):=\{h:B\rightarrow C\mid h\circ f\in S\}=
\{h:B\rightarrow C\}=:\ \uparrow\!\!B \label{f*S}
\end{equation}
where $\uparrow\!\!B$ denotes the {\em principal\/} sieve on
$B$, defined to be the set of all arrows in $\cal C$ whose
domain is $B$.

A crucial property of sieves is that the set ${\bf\Omega}(A)$
of sieves on $A$ has the structure of a Heyting algebra where
the unit element $1_{{\bf\Omega}(A)}$ in ${\bf\Omega}(A)$ is
the principal sieve $\uparrow\!\!A$, and the null element
$0_{{\bf\Omega}(A)}$ is the empty sieve $\emptyset$. The
partial ordering in ${\bf\Omega}(A)$ is defined by $S_1\leq
S_2$ if and only if $S_1\subseteq S_2$; and the logical
connectives are defined as:
\begin{eqnarray}
    && S_1\land S_2:=S_1\cap S_2    \label{Def:S1landS2}\\
    && S_1\lor S_2:=S_1\cup S_2     \label{Def:S1lorS2} \\
    &&S_1\Rightarrow S_2:=\{f:A\rightarrow B\mid
    \forall g:B\rightarrow C {\rm\ if\ } g\circ f\in S_1
        {\rm\ then\ }g\circ f\in S_2\}\hspace{1cm}.
\end{eqnarray}
As in any Heyting algebra, the negation of an element $S$
(called the {\em pseudo-complement\/} of $S$) is defined as
$\neg S:=S\Rightarrow 0$; so that
\begin{equation}
    \neg S:=\{f:A\rightarrow B\mid \mbox{for all
$g:B\rightarrow C$, $g\circ f\not\in S$} \}. \label{Def:negS}
\end{equation}
As remarked earlier, the main distinction between a Heyting
algebra and a Boolean algebra is that, in the former, the
negation operation does not necessarily obey the law of
excluded middle: instead, all that be can said is that, for
any element $S$,
\begin{equation}
        S\lor\neg S\leq 1.
\end{equation}

It can be shown that the presheaf ${\bf\Omega}$ is a subobject
classifier for the topos ${\rm Set}^{{\cal C}}$. Thus subobjects
of any object $\bf X$ in this topos ({\em i.e.}, any presheaf on
$\cal C$) are in one-to-one correspondence with arrows $\chi:{\bf
X}\rightarrow {\bf\Omega}$. This works as follows. Let $\bf K$ be
a subobject of $\bf X$. Then there is an associated {\em
characteristic\/} arrow $\chi^{{\bf K}}:{\bf
X}\rightarrow{\bf\Omega}$, whose component $\chi^{{\bf K}}_A:{\bf
X}(A)\rightarrow{\bf\Omega}(A)$ at each `stage of truth' $A$ in
$\cal C$ is defined as
\begin{equation}
    \chi^{{\bf K}}_A(x):=\{f:A\rightarrow B\mid {\bf X}(f)(x)\in
{\bf K}(B)\} \label{Def:chiKA}
\end{equation}
for all $x\in {\bf X}(A)$. That the right hand side of Eq.\
(\ref{Def:chiKA}) actually {\em is\/} a sieve on $A$ follows
from the defining properties of a subobject.\footnote{There is
a converse to Eq.\ (\ref{Def:chiKA}): namely, each arrow
$\chi:{\bf X}\rightarrow{\bf\Omega}$ ({\em i.e.}, a natural
transformation between the presheaves ${\bf X}$ and
${\bf\Omega}$) defines a subobject ${\bf K}^\chi$ of $\bf X$
via
\begin{equation}
    {\bf K}^\chi(A):=\chi_A^{-1}\{1_{{\bf\Omega}(A)}\}.
                            \label{Def:KchiA}
\end{equation}
at each stage of truth $A$.}

Thus, in each `branch' of the category $\cal C$ going
`upstream' from the stage $A$, $\chi^{{\bf K}}_A(x)$ picks out
the first member $B$ in that branch for which ${\bf X}(f)(x)$
lies in the subset ${\bf K}(B)$, and the commutative diagram
Eq.\ (\ref{subobject}) then guarantees that ${\bf X}(h\circ
f)(x)$ will lie in ${\bf K}(C)$ for all $h:B\rightarrow C$.

Thus each `stage of truth' $A$ in $\cal C$ serves as a possible
{\em context\/} for an assignment to each $x\in {\bf X}(A)$ of a
generalised truth-value, which is a sieve belonging to the Heyting
algebra ${\bf\Omega}(A)$. This is the sense in which contextual,
generalised truth-values arise naturally in a topos of presheaves.

\subsubsection{Global Sections of a Presheaf}
For the category of presheaves on $\cal C$, a terminal object
${\bf 1}:{\cal C}\rightarrow {\rm Set}$ can be defined by ${\bf
1}(A):=\{*\}$ (a singleton set) at all stages $A$ in $\cal C$; if
$f:A\rightarrow B$ is an arrow in $\cal C$ then ${\bf
1}(f):\{*\}\rightarrow\{*\}$ is defined to be the map $*\mapsto
*$. This is indeed a terminal object since, for any presheaf $\bf
X$, we can define a unique natural transformation $N:{\bf
X}\rightarrow{\bf 1}$ whose components $N_A:{\bf
X}(A)\rightarrow{\bf 1}(A)=\{*\}$ are the constant maps $x\mapsto
*$ for all $x\in{\bf X}(A)$.

A global element (or point) of a presheaf $\bf X$ is also
called a {\em global section\/}. As an arrow $\gamma:{\bf
1}\rightarrow{\bf X}$ in the topos ${\rm Set}^{{\cal C}}$, a
global section corresponds to a choice of an element
$\gamma_A\in{\bf X}(A)$ for each stage of truth $A$ in $\cal
C$, such that, if $f:A\rightarrow B$, the `matching condition'
\begin{equation}
    {\bf X}(f)(\gamma_A)=\gamma_B \label{Def:global}
\end{equation}
is satisfied. As we shall see, the Kochen-Specher theorem can
be read as asserting the non-existence of any global sections
of certain presheaves that arises naturally in quantum theory.

\section{ Presheaves of Propositions, and Valuations in Quantum
Theory} \label{Sec:PreProValQuaThe} The contextual,
multi-valued logic that arises naturally in a topos of
presheaves has some very interesting potential applications in
theoretical physics. Here, however, I shall briefly present
just one particular example that has been developed in detail
elsewhere \cite{IB98,IB99,HIB99}. This is the proposal to
retain a `realist flavour' in the assignment of values to
quantum-theoretic quantities by using the non-Boolean logical
structure of a particular topos of presheaves.

Before stating the proposal precisely,  recall the
Kochen-Specher theorem which asserts the impossibility of
associating real values $V(\hat A)$ to all physical quantities
in a quantum theory (if $\dim {\cal H}>2$) whilst preserving
the `{\em FUNC\/}' rule that $V(f(\hat A)=f(V(\hat A))$---{\em
i.e.}, the value of a function $f$ of a physical quantity $A$
is equal to the function of the value of the quantity.
Equivalently, it is not possible to assign true-false values
to all the propositions in a quantum theory in a way that
respects the structure of the associated lattice of projection
operators. As we shall see, our topos-theoretic proposal is
such that  the truth value ascribed to a proposition about the
value of a physical quantity need not be just `true' or
`false'.

Thus consider the proposition ``$A \in \Delta$'', which
asserts that the value of the quantity $A$ lies in a Borel set
$\Delta \subseteq \mathR $. Roughly speaking, our proposal is
that any such proposition should be ascribed as a truth-value
a set of coarse-grainings, $f(\hat{A})$, of the operator
$\hat{A}$ that represents $A$. Exactly which coarse-grainings
are in the truth-value depends in a precise way on $\Delta$
and the quantum state $\psi$: specifically, $f(\hat{A})$ is in
the truth-value if and only if $\psi$ is in the range of the
spectral projector $\hat{E}[f(A) \in f(\Delta)]$. Note the
contrast with the conventional eigenstate-eigenvalue link: our
requirement is not that $\psi$ be in the range of $\hat{E}[A
\in \Delta]$, but a weaker one since, generally, $\hat{E}[f(A)
\in f(\Delta)]$ is a larger spectral projector than
$\hat{E}[f(A) \in f(\Delta)]$; {\em i.e.\/}, in the lattice
${\cal L}({\cal H})$ of projectors on the Hilbert space $\cal
H$, we have $\hat{E}[A \in \Delta] \leq \hat{E}[f(A) \in
f(\Delta)]$.

 So the intuitive idea is that the new proposed truth-value of ``$A
\in \Delta$'' is given by the set of weaker propositions
``$f(A) \in f(\Delta)$'' that are true in the old ({\em i.e.},
eigenstate-eigenvalue link) sense. More precisely, the
truth-value of ``$A \in \Delta$'' is the set of quantities
$f(A)$ for which the corresponding weaker proposition ``$f(A)
\in f(\Delta)$'' is true in the old sense. Thus the
truth-value of a proposition in the new sense is given by the
set of its consequences that are true in the old sense.

The first step in stating the proposal precisely is to
introduce the set ${\cal O}$ of all bounded self-adjoint
operators on the Hilbert space $\cal H$ of a quantum system.
The set ${\cal O}$ is turned into a category by defining the
objects to be the elements of ${\cal O}$, and saying that
there is an arrow from $\hat A$ to $\hat B$ if there exists a
real-valued function $f$ on the spectrum $\sigma(\hat
A)\subset\mathR$ of $\hat A$, such that $\hat B=f(\hat A)$. If
$\hat B=f(\hat A)$, for some $f:\sigma(\hat
A)\rightarrow\mathR$, then the corresponding arrow in the
category ${\cal O}$ will be denoted $f_{{\cal O}}: \hat
A\rightarrow\hat B$.

The next step is define two presheaves on the category ${\cal
O}$, called the {\em dual presheaf\/} and the {\em
coarse-graining presheaf\/} respectively. The former affords
an elegant formulation of the Kochen-Specker theorem, namely
as the statement that the dual presheaf does not have global
sections. The latter is at the basis of the proposed
generalised truth-value assignments.

The dual presheaf on ${{\cal O}}$ is the covariant functor
${\bf D}:{\cal O}\rightarrow {\rm Set}$ defined as follows:
\begin{enumerate}
    \item On objects: ${\bf D}(\hat A)$ is the {\em dual\/} of
$W_A$, where $W_A$ is the spectral algebra of the operator
$\hat A$ ({\em i.e.} $W_A$ is the collection of all projectors
onto the subspaces of $\cal H$ associated with Borel subsets
of $\sigma(\hat A)$). Thus ${\bf D}(\hat A)$ is defined to be
the set of all homomorphisms from the Boolean algebra $W_A$ to
the Boolean algebra $\{0,1\}$.

    \item On arrows: If $f_{{\cal O}}:\hat A\rightarrow \hat B$,
so that $\hat B=f(\hat A)$, then ${\bf D}(f_{\cal
O}):D(W_A)\rightarrow D(W_B)$ is defined by ${\bf D}(f_{\cal
O})(\chi):= \chi|_{W_{f(A)}}$ where $\chi|_{W_{f(A)}}$ is the
restriction of $\chi\in D(W_A)$ to the subalgebra
$W_{f(A)}\subseteq W_A$.
\end{enumerate}

    A global element (global section) of the functor ${\bf
D}:{\cal O}\rightarrow {\rm Set}$ is then a function $\gamma$
that associates to each ${\hat A} \in\cal O$ an element
$\gamma_{A}$ of the dual of $W_A$ such that if $f_{\cal
O}:{\hat A}\rightarrow {\hat B}$ (so $\hat B=f(\hat A)$ and
$W_{B}\subseteq W_A$), then $\gamma_{A}|_{W_B}=\gamma_{B}$.
Thus, for all projectors $\hat\alpha\in W_{B}\subseteq W_A$,
we have $\gamma_{B}(\hat\alpha)=\gamma_{A}(\hat\alpha)$.

Since each $\hat\alpha$ in the lattice $\cal {L(H)}$ of
projection operators on $\cal H$ belongs to at least one such
spectral algebra $W_A$ (for example, the algebra $\{\hat
0,\hat 1,\hat\alpha,\hat 1 - \hat\alpha\}$) it follows that a
global section of ${\bf D}$ associates to each projection
operator $\hat\alpha \in \cal {L(H)}$ a number $V(\hat\alpha)$
which is either $0$ or $1$, and is such that if $\hat\alpha$
and $\hat\beta$ are disjoint propositions then
$V(\hat\alpha\lor\hat\beta)=V(\hat\alpha)+V(\hat\beta)$. A
global section $\gamma$ of the presheaf ${\bf D}$ would
correspond to an assignment of truth-values $\{0,1\}$ to all
propositions of the form ``$A\in\Delta$'', which obeyed the
{\em FUNC\/} condition $\gamma_{A}|_{W_B}=\gamma_{B}$. But
these are precisely the types of valuation prohibited by the
Kochen-Specker theorem provided that $\dim{\cal H}>2$! So an
alternative way of expressing the Kochen-Specker theorem is
the statement that (if $\dim{\cal H}>2$) the dual presheaf
${\bf D}$ has no global sections.

However, we {\em can} use the subobject classifier $\bf \Omega$ in
the topos ${\rm Set}^{\cal O}$ of all presheaves on $\cal O$ to
assign {\em generalized} truth-values to the propositions
``$A\in\Delta$''. These truth-values will be sieves---as defined
in Section \ref{SubSub:Sieves}---and since they will be assigned
relative to each `context' or `stage of truth' $\hat A$ in $\cal
O$, these truth-values will be contextual as well as generalized.
Note that because in any topos the subobject classifier $\bf
\Omega$ is unique up to isomorphism the traditional objection to
multi-valued logics in quantum theory---that their structure often
seems arbitrary---does not apply to these particular generalized,
contextual truth-values.

The first step is to define the appropriate presheaf of
propositions. The {\em coarse-graining presheaf\/} over $\cal
O$ is the covariant functor ${\bf G}:{\cal O}\rightarrow{\rm
Set}$ defined as follows.
\begin{enumerate}
\item {\em On objects in $\cal O$:} ${\bf G}(\hat A):=
W_A$, the spectral algebra of $\hat A$.

\item {\em On arrows in $\cal O$:} If $f_{\cal O}:\hat A
\rightarrow\hat B$ ({\em i.e.}, $\hat B=f(\hat A)$), then
${\bf G}(f_{\cal O}): W_A\rightarrow W_B$ is defined
as\footnote{If $f(\Delta)$ is not Borel, the right hand side
is to be understood in the sense of Theorem 4.1 of
\cite{IB98}---a measure-theoretic nicety that we shall not
discuss here.}
\begin{equation}
        {\bf G}(f_{\cal O})(\hat E[A\in\Delta]):=
        \hat E[f(A)\in f(\Delta)]       \label{Def:G(fO)}
\end{equation}
\end{enumerate}

A function $\nu$ that assigns to each object $\hat A$ in $\cal
O$ and each Borel set $\Delta \subseteq \sigma(\hat A)$, a
sieve of arrows in $\cal O$ on $\hat A$ ({\em i.e.}, a sieve
of arrows with $\hat A$ as domain), will be called a {\em
sieve-valued valuation\/} on $\bf G$. We write the values of
this function as $\nu(A \in \Delta)$.

From the logical point of view, a natural requirement for any
kind of valuation on a presheaf of propositions such as $\bf
G$ is that the valuation should specify a subobject of $\bf
G$.  But subobjects are in one-one correspondence with arrows,
{\em i.e.}, natural transformations, $N:{\bf G}\rightarrow
{\bf\Omega}$. So it is natural to require a sieve-valued
valuation $\nu$ to define such a natural transformation by the
equation $N^\nu_A(\hat E[A\in\Delta]):=\nu(A \in \Delta)$ for
all stages/contexts $\hat A$.

This requirement leads directly to the analogue for presheaves
of the functional composition condition of the Kochen-Specker
theorem, called {\em FUNC\/} above. Indeed, it transpires that
a sieve-valued valuation defines a natural transformation if
and only if it obeys (the presheaf version of) {\em FUNC\/}.

To spell this out, first recall that sieves are `pushed forward'
by the subobject classifier $\bf \Omega$ according to Eq.\
(\ref{Def:Om(f)}). For the category $\cal O$: if $f_{\cal O}:\hat
A\rightarrow\hat B$, then ${\bf\Omega}(f_{\cal
O}):{\bf\Omega}(\hat A)\rightarrow{\bf\Omega}(\hat B)$ is defined
by
\begin{equation}
{\bf\Omega}(f_{\cal O})(S):= \{h_{\cal O}:B\rightarrow C\mid
h_{\cal O}\circ f_{\cal O}\in S\}
                                \label{Def:Om(f)forO}
\end{equation}
for all sieves $S\in{\bf\Omega}(\hat A)$.

Accordingly, we say that a sieve-valued valuation $\nu$ on
$\bf G$ satisfies {\em generalized functional
composition\/}---for short, {\em FUNC\/}---if for all $\hat
A,\hat B$ and $f_{\cal O}:\hat A\rightarrow\hat B$ and all
${\hat E}[A \in \Delta]\in{\bf G}(\hat A)$, we have
\begin{equation}
    \nu(B \in {\bf G}(f_{\cal O})({\hat E}[A \in \Delta])) \equiv \nu(f(A)
\in f(\Delta)) = {\bf\Omega}(f_{\cal O})(\nu(A \in \Delta)).
\label{Def:GenlzdFunc}
\end{equation}

It can readily be  checked that {\em FUNC\/} is exactly the
condition a sieve-valued valuation must obey in order to
define a natural transformation---{\em i.e.}, a subobject of
$\bf G$---by the equation $N^\nu_A(\hat E[A\in\Delta]):=\nu(A
\in \Delta)$. That is, a sieve-valued valuation $\nu$ on $\bf
G$ obeys {\em FUNC\/} if and only if the functions at each
context $\hat A$
\begin{equation}
    N^\nu_{\hat A}(\hat E[A\in\Delta]):=\nu(A \in \Delta)
\end{equation}
define a natural transformation $N^\nu$ from ${\bf G}$ to
${\bf\Omega}$.

It turns out that with any quantum state there is associated such
a sieve-valued valuation obeying {\em FUNC\/}. Furthermore, this
valuation gives the natural generalization of the
eigenvalue-eigenstate link described earlier. That is, a quantum
state $\psi$ induces a sieve on each $\hat A$ in $\cal O$ by the
requirement that an arrow $f_{\cal O}:\hat A\rightarrow\hat B$ is
in the sieve if and only if $\psi$ is in the range of the spectral
projector $\hat{E}[B \in f(\Delta)]$. To be precise, we define for
any $\psi$, and any Borel subset $\Delta$ of the spectrum
$\sigma(\hat A)$ of $\hat A$,
\begin{eqnarray}
    \nu^\psi(A\in\Delta)&:=&\{f_{\cal O}:\hat A\rightarrow\hat B
        \mid \hat E[B\in f(\Delta)]\psi=\psi\} \nonumber\\[2pt]
    &=&  \{f_{\cal O}:\hat A\rightarrow\hat B
        \mid {\rm Prob}(B\in f(\Delta);\psi)=1\}
                \label{Def:nupsiDelta}
\end{eqnarray}
where ${\rm Prob}(B\in f(\Delta);\psi)$ is the usual Born-rule
probability that the result of a measurement of $B$ will lie
in $f(\Delta)$, given the state $\psi$.

One can check that the definition satisfies {\em FUNC\/}, and also
has other properties that it is natural to require of a valuation
(discussed in \cite{IB98,IB99,HIB99}). Thus, by using topos theory
we are able to assign generalised truth values to all propositions
whilst preserving the appropriate analogue of the FUNC condition.

The key feature of these truth assignments is that they
involve the contextual, multi-valued logic that is an
intrinsic feature of a topos of presheaves. My expectation is
that a similar topos structure could serve to patch together
the different types of quantum theory that, as discussed
earlier, I anticipate should be associated with different
background space-time structures.

\section{Conclusions}
In general, the real numbers enter physical theories in three
ways: as the values of physical quantities; as coordinates on a
manifold model for space and time; and as the values of
probabilities. The main thrust of the present paper is to argue
that all three uses may become problematic in physical regimes
that characterise strong quantum gravity effects.

In particular, I have argued that the assignment of real
numbers as values of physical quantities and probabilities is
to some extent motivated by certain {\em a priori\/} ideas
about the continuum nature of space and time. Thus it may be
fundamentally wrong to attempt to construct a quantum theory
of gravity whilst using a quantum formalism in which these
{\em a priori\/} continuum ideas are present from the
beginning. My contention is that there should be a different
{\em type\/} of quantum structure for each `type' of
background space-time: in particular, the mathematical spaces
in which physical quantities and probabilities take their
values should reflect the structure of this background.

If this is correct, the question then arises as to how to
patch together a collection of such theories in the situation
where the `background' space-times are themselves the subject
of quantum effects. I have suggested that the appropriate
mathematical tool for doing this is topos theory; in
particular the theory of presheaves with its intrinsic
contextual, multi-valued logic. As an example of the use of
this theory I have briefly reviewed an application of presheaf
theory to the Kocken-Specher theorem in standard quantum
theory.

Topos theory is also an essential ingredient in synthetic
differential geometry, and this too may have important
applications in theoretical physics; particularly perhaps in
the context of the two-pronged way in which time arises in the
consistent history theory.

What is sketched in the first half of this paper is only a
collection of ideas. It remains an outstanding challenge to
implement some of these general thoughts in the context, say,
of a specific non-standard spatio-temporal background, such as
a causal set. This could give valuable insight into what is
perhaps the hardest task of all: to construct a quantum
formalism for use in situations where there are no {\em prima
facie\/} spatio-temporal concepts at all---a situation that
could well arise in a quantum gravity theory in which all of
what we might want to call ``spatio-temporal concepts'' emerge
from the basic formalism only in some limiting sense.

\bigskip\noindent
{\bf Acknowledgements} Some of the material presented here is
a development of earlier work  \cite{IB00} with Jeremy
Butterfield. I am grateful to him for permission to include
this material. I am also very grateful to Ntina Savvidou for
detailed discussions about the present paper. Support by the
EPSRC in form of grant GR/R36572 is gratefully acknowledged.


\begin{thebibliography}{10}
\bibitem{IB00}
C.J.~Isham and J.~Butterfield.
\newblock Some possible roles for topos theory in quantum theory
and quantum gravity.
\newblock {\em Found.\ Phys.}, 30:1707--1735, (2000).

\bibitem{BMLS87}
L.~Bombelli, J.~Lee, D.~Meyer and R.D.~Sorkin.
\newblock Spacetime as a causal set.
\newblock {\em Phys.\ Rev.\ Lett.}, 30:521--524, (1987).

\bibitem{Sorkin91} R.D.~Sorkin.
\newblock Spacetime and Causal Sets.
\newblock In J.C. D'Olivo, E. Nahmad-Achar, M. Rosenbaum,
M.P. Ryan, L.F. Urrutia and F. Zertuche (eds.),
   {\em Relativity and Gravitation:  Classical and Quantum,}
\newblock(Proceedings of the {\it SILARG VII Conference},
    held Cocoyoc, Mexico, December, 1990), pages 150--173,
\newblock World Scientific, Singapore, (1991).

\bibitem{isham-rev2} C.~J.~Isham.
\newblock Canonical quantum gravity and the problem
     of time.
\newblock In {\em Integrable Systems, Quantum Groups, and Quantum
Field Theories}, eds. L.~A.~Ibort and M.~A.~Rodriguez, Kluwer
Academic Publishers, London, pp 157--288 (1993).

\bibitem{kuch-prehled} K.~V.~Kucha\v{r}.
\newblock Time and interpretations of quantum gravity.
\newblock In {\em Proceedings of the 4th Canadian
    Conference on General Relativity and Relativistic Astrophysics},
    World Scientific, Singapore, pp 211--314 (1992).

\bibitem{Gri84} R.B. Griffiths.
\newblock Consistent histories and the interpretation of quantum mechanics.
\newblock {\em J. Stat. Phys.}, 36:219--272, (1984).

\bibitem{Omn88} R.~Omn\`es.
\newblock Logical reformulation of quantum mechanics. {I.} {F}oundations.
\newblock {\em J. Stat. Phys.}, 53:893--932, (1988).

\bibitem{GH90} M.~Gell-{M}ann and J.~Hartle.
\newblock Quantum mechanics in the light of quantum cosmology.
\newblock In W.~Zurek, editor, {\em Complexity, Entropy and the Physics of
  Information, SFI Studies in the Science of Complexity, {Vol. VIII}}, pages
  425--458. Addison-Wesley, Reading, (1990).

\bibitem{Isham94} C.J. Isham.
\newblock Quantum  logic and the histories approach to quantum theory.
\newblock {\em J. Math. Phys. } 35:2157--2185, (1994).

\bibitem{IL94}
C.J. Isham and N.~Linden.
\newblock Quantum temporal logic and decoherence functionals in the histories
approach to generalised quantum theory.
\newblock {\em J. Math. Phys. } 35:5452--5476, (1994).

\bibitem{Far75} M.~Farrukh.
\newblock Application of nonstandard analysis to quantum mechanics.
\newblock {\em J.\ Math.\ Physics} 16:177--200, (1975).

\bibitem{Lav96} R.~Lavendhomme.
\newblock {\em Basic Concepts of Synthetic Differential
Geometry}.
\newblock Kluwer, Dordrecht, (1996).

\bibitem{Fearns02} J.~Fearns.
\newblock A physical quantum model in a smooth topos.
\newblock quant-ph/0202079, (2002).

\bibitem{Nishimura97} H.~Nishimura.
\newblock Synthetic hamiltonian mechanics.
\newblock {\em Int.\ Jour.\ Theor.\ Phys.} 36:259--279 (1997)

\bibitem{ILSS98}
\newblock  C.~Isham, N.~Linden, K.~Savvidou and S.~Schreckenberg.
\newblock  Continuous time and consistent histories.
\newblock  {\em J.\ Math.\ Phys.} 39:1818--1834, (1998).

\bibitem{Sav99a} K.~Savvidou.
\newblock The action operator for continuous-time histories.
\newblock  {\em J.\ Math.\ Phys.}  40: 5657, (1999).

\bibitem{Sav99b} K.~Savvidou.
\newblock  Continuous Time in Consistent Histories.
\newblock  {PhD Thesis} in gr-qc/9912076, (1999).

\bibitem{SavGR1} K.~Savvidou.
\newblock General relativity histories theory.
\newblock {\em Class.\ Quant.\ Grav.} 18: 3611-3628, (2001).

\bibitem{Gol84}
R.~Goldblatt.
\newblock {\em Topoi: The Categorial Analysis of Logic}.
\newblock North-Holland, London, (1984).

\bibitem{MM92}
S.~Mac{L}ane and I.~Moerdijk.
\newblock {\em Sheaves in Geometry and Logic: {A} First
Introduction to Topos Theory}.
\newblock Springer-Verlag, London, (1992).

\bibitem{IB98}
C.J. Isham and J.~Butterfield.
\newblock A topos perspective on the {K}ochen-{S}pecker theorem:
{I.} {Q}uantum states as generalised valuations.
\newblock {\em Int.\ J.\ Theor.\ Phys.}, 37: 2669--2733, (1998).

\bibitem{IB99}
J.~Butterfield and C.J.~Isham.
\newblock A topos perspective on the {K}ochen-{S}pecker theorem:
{II.} {C}onceptual aspects, and classical analogues.
\newblock {\em Int.\ J.\ Theor\ Phys.}, 38: 827--859, (1999).

\bibitem{HIB99}
J.~Hamilton, C.J.~Isham and J.~Butterfield.
\newblock A topos perspective on the {K}ochen-{S}pecker theorem:
{III.} Von Neumann algebras as the base category.
\newblock {\em Int.\ J.\ Theor\ Phys.}, 38: 827--859, (2000).

\bibitem{KS67}
S.~Kochen, and E.~Specker.
\newblock The problem of hidden variables in quantum mechanics.
\newblock {\em J.\ Math.\ and Mech.} 39: 59--87, (1967).

\end{thebibliography}
\end{document}